\documentclass[twocolumn]{aastex62}
\usepackage{graphicx}
\usepackage[normalem]{ulem}
\newcommand\redout{\bgroup\markoverwith
{\textcolor{red}{\rule[.5ex]{2pt}{0.4pt}}}\ULon}
\usepackage{xcolor}
\usepackage{threeparttable}
\usepackage{amsmath}

\newcommand{\jqso}{J120110.31+211758.5}
\newcommand{\zdla}{3.7975}
\newcommand{\dlaname}{DLA1201+2117\_3.80}

\newcommand{\origpos}{J120110.26+211756.2}
\newcommand{\cplus}{[\ctwo]~158\,$\mu$m}
\newcommand{\tsiglim}{0.18}  
\newcommand{\zcube}{3.7978}

\newcommand{\mnhi}{N_{\rm HI}}
\newcommand{\nhi}{$\mnhi$}
\newcommand{\cmt}{{\rm cm}^{-2}}

\newcommand{\newcyc}{5}      
\newcommand{\bwidth}{0\farcs33}    
\newcommand{\spatres}{2.3}    

\newcommand{\mrperp}{R}

\newcommand{\vrperp}{18}  
\newcommand{\rsep}{6}  

\newcommand{\mdvninety}{\Delta v_{\rm 90}}
\newcommand{\dvninety}{$\mdvninety$}
\newcommand{\mdvmax}{\mdvninety^{\rm max}}
\newcommand{\dvmax}{$\mdvmax$}
\newcommand{\mfdv}{f(\mdvninety)}
\newcommand{\fdv}{$\mfdv$}
\newcommand{\dvntwo}{200}    
\newcommand{\dvnfour}{260}    

\newcommand{\mfmdla}{f_m^{\rm DLA}}    
\newcommand{\fmdla}{$\mfmdla$}
\def\intl{\int\limits}
\newcommand{\madla}{A_{\rm DLA}}  
\newcommand{\adla}{$\madla$}
\newcommand{\mfmg}{f_{MM}^{\rm g}}    
\newcommand{\fmg}{$\mfmg$}

\newcommand{\aslope}{-1.0}    
\newcommand{\ayoff}{+1.0\,kpc}     
\newcommand{\bslope}{-1.0}    
\newcommand{\byoff}{-2.0\,kpc}     

\newcommand{\mldust}{L_{\rm IR}}  
\newcommand{\ldust}{$\mldust$}
\newcommand{\mlcplus}{L_{\rm C^+}}   
\newcommand{\lcplus}{$\mlcplus$}

\newcommand{\mkms}{\rm km \, s^{-1}}

\newcommand{\fetwo}{Fe\textsc{ii}}

\newcommand{\mstareq}{M_*}

\newcommand{\mmsun}{M_{\odot}}

\newcommand{\hi}{H\textsc{i}}

\newcommand{\ctwo}{C\textsc{ii}}

\newcommand{\lya}{Ly$\alpha$}

\newcommand{\beq}{\begin{equation}}
\newcommand{\eeq}{\end{equation}}

\graphicspath{{./}{figures/}}

\submitjournal{ApJL}

\shorttitle{ALMA C{\sc ii} 158$\mu$m imaging of an H{\textsc i}-Selected Major Merger at $z \sim 4$}
\shortauthors{Prochaska et al.}

\begin{document}


\title{ALMA C{\sc ii} 158$\mu$m Imaging of an H{\textsc i}-Selected Major Merger at $z \sim 4$}

\correspondingauthor{J. Xavier Prochaska}
\email{xavier@ucolick.org}

\author[0000-0002-7738-6875]{J. Xavier Prochaska}
\affiliation{Department of Astronomy \& Astrophysics, UCO/Lick Observatory, University of California, 1156 High Street, Santa Cruz, CA 95064, USA}
\affiliation{
Kavli Institute for the Physics and Mathematics of the Universe (Kavli IPMU),
5-1-5 Kashiwanoha, Kashiwa, 277-8583, Japan}
\author[0000-0002-9838-8191]{Marcel Neeleman}
\affiliation{Max-Planck-Institut f\"{u}r Astronomie, K\"{o}nigstuhl 17, D-69117, Heidelberg, Germany}
\author[0000-0002-9757-7206]{Nissim Kanekar}
\altaffiliation{Swarnajayanti Fellow}
\affiliation{National Centre for Radio Astrophysics, Tata Institute of Fundamental Research, Pune University, Pune 411007, India}
\author[0000-0002-9946-4731]{Marc Rafelski}
\affiliation{Space Telescope Science Institute, Baltimore, MD 21218, USA}
\affiliation{Department of Physics \& Astronomy, Johns Hopkins University, Baltimore, MD 21218, USA}



\begin{abstract}

We present high spatial-resolution ($\approx 2$\,kpc) Atacama Large Millimeter/submillimeter 
Array (ALMA) observations of \cplus\ and dust-continuum emission from a galaxy at $z=\zcube$ 
selected by its strong \hi\ absorption (a damped \lya\ absorber, DLA) against a background QSO. 
Our ALMA images reveal a pair of star-forming galaxies separated by $\approx \rsep$\,kpc 
(projected) undergoing a major merger. Between these galaxies is a third emission component 
with highly elevated ($2\times$)
\cplus\ emission relative to the dust continuum, which is likely to arise 
from stripped gas associated with the merger. This merger of two otherwise-normal galaxies 
is not accompanied by enhanced star-formation, contrary to mergers detected in most 
luminosity-selected samples. The DLA associated with the merger exhibits extreme kinematics, 
with a velocity width for the low-ionization metal lines of $\mdvninety \approx 470 \, \mkms$, that 
spans the velocity spread revealed in the \cplus\ emission. We propose that DLAs with high 
\dvninety\ values are a signpost of major mergers in normal galaxies at high redshifts, and 
use the distribution of the velocity widths of metal lines in high-$z$ DLAs to provide
a rough estimate the 
fraction of $z > 3$ galaxies that are undergoing a major merger.
\end{abstract}




\section{Introduction}

At the heart of modern cosmology is the concept that large structures are built up 
from the collapse and agglomeration of smaller structures. This holds for both dark 
matter halos and the galaxies within them, which together are predicted to merge 
across cosmic time to yield the massive galaxies observed today. Such 
mergers are observed and predicted to transform the morphology of galaxies 
\citep[e.g.][]{mh96,naab2003,lotz+08b}, to dominate the mass evolution in quiescent 
galaxies \citep[e.g.][]{bell+06}, and to trigger dramatic enhancements or declines
in a galaxy's star formation rate \citep[e.g.][]{spf01,jogee+2009,patton+2011}.
One is thus motivated to track the incidence and processes that govern galaxy mergers 
across cosmic time.

Despite these strong motivations to study mergers, empirical studies -- especially in the high-$z$ 
universe -- are relatively scarce. Indeed, the majority of merging systems known
are serendipitous discoveries \citep[e.g.][]{cooke+2010} or are associated with 
the most luminous galaxies, e.g.\ sub-mm galaxies with 
extreme star-formation rates \citep[SFRs;][]{olivares+16}, or galaxies hosting 
ultra-luminous quasars \citep{neeleman+19b}.
There is also a set of imaging studies that statistically identify galaxies likely to
merge within approximately a dynamical time \citep[e.g.][]{patton+1997,bundy+2009,mantha+2018}.
These studies are limited by the lack of spectroscopy which means ambiguous identifications
of individual mergers and, more importantly, no analysis of the system dynamics
\citep[e.g.][]{ellison+10}.

An important exception are the analyses by \citet{ventou+2017,ventou+2019}
who used deep VLT/MUSE integral field spectroscopy to study mergers of sub-$L^*$ galaxies. 
These studies are, however, flux-limited and therefore prone to selection biases.
Such biases could manifest in either direction, e.g.\ a higher inferred merger rate if 
gas is preferentially driven towards the central regions to ignite star-formation 
and/or AGN activity \citep[e.g.][]{mh96}, or a lower rate if processes driving the 
merger enshroud the system in dust and obscure it from detection \citep[e.g.][]{treister+10}.

An additional tenant of our cosmological paradigm is that the rate of assembly on 
galaxy scales was greater in the past. Analyses of cosmological simulations report that on average major mergers (defined here as at least a 1:4 mass ratio) occur every $\approx 2$~Gyr at $z=3$ for galaxies with 
stellar mass $\mstareq \gtrsim 10^{11} \mmsun$ \citep{aldo+2017}.
The merger rate is expected to increase with redshift, with 
mergers playing an increasingly important role in galaxy assembly at early times.

\begin{figure*}
\centering
\includegraphics[width=\textwidth]{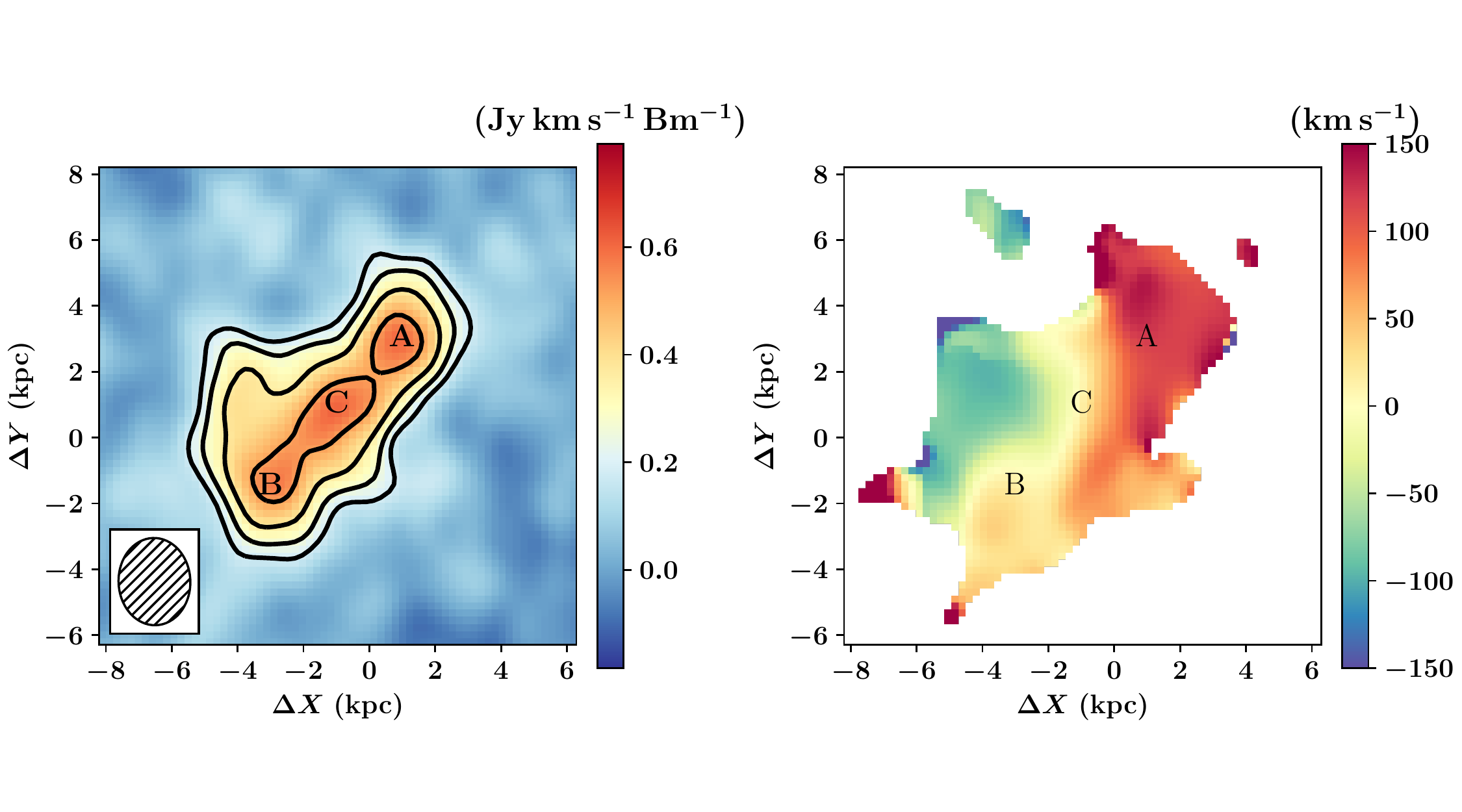}
\vspace{-0.3in}
\caption{(left) The zeroth moment image showing the integrated \cplus\ flux density 
of the galaxies A and B associated with \dlaname, here centered on \origpos.
The axes are labeled in physical units (kpc) at $z=3.7978$ and with 
$\Delta X$, the negative offset in RA, and $\Delta Y$, the offset in Dec, relative 
to \origpos. 
The lowest contour corresponds to $3\sigma$ significance 
(\tsiglim~Jy km s$^{-1}$ beam$^{-1}$), with the contours increasing by a factor 
of 2$^{1/2}$. 
(right) The moment-1 image showing the flux density-weighted velocity 
field, restricted to regions where the integrated flux density exceeds 2.5$\sigma$ 
significance. There are two kinematically distinct components which we associate 
with a pair of merging galaxies, labelled A and B. There is additional emission between 
these components, referred to as component~C (see text).
}
\label{fig:moments}
\end{figure*}

Empirically, it is challenging to assess the merger rate because we cannot 
directly estimate the timescale for these events from the `static' observations 
afforded on human timescales.  Instead, the majority of studies have focused on 
the fraction of galaxies in a population that are interacting and/or likely to 
undergo a major merger, \fmg.  \fmg\ is frequently defined as the fraction 
of galaxies with projected physical separation $\delta R \le 25$\,kpc and velocity 
offset $\delta v < 500 \, \mkms$. The current literature reports a low major merger 
fraction today \citep[$\mfmg \approx 0.01$][]{xu+2012}, with a rise from $z=0$ to 
$z\approx 2$, peaking at $\mfmg \approx 2$ \citep{lopez+2013,ventou+2017,ventou+2019}.  
At high redshifts ($z> 2$), the results are drawn solely from the VLT/MUSE surveys of 
\cite{ventou+2017,ventou+2019}, which report a surprising decline in \fmg\ at $z > 2$, 
in apparent contradiction with the theoretical rise in the accretion rate. This apparent 
conundrum has been resolved by recognizing that the time to merge is much shorter in the 
high density environment of the high-$z$ Universe.
Interpretations aside, 
this provides a limited dataset to examine the merger fraction and/or the detailed 
impacts of the merging process on high-$z$ galaxies.

Over the last few years, we have used the Atacama Large Millimeter/submillimeter Array (ALMA) to search for redshifted \cplus\ emission from a sample of \hi-selected galaxies at $z \gtrsim 4$ \citep{neeleman+17,neeleman+19}. These galaxies were targeted because of their association with a high column density of \hi\ (\nhi), identified as a damped  \lya\ absorber \citep[DLA;][]{wgp05} in a background quasar spectrum. 
Given the paucity of quiescent galaxies devoid of an interstellar medium at $z>2$,  
\hi-selection should yield galaxies spanning nearly the complete distribution at any redshift. 
Our ALMA \cplus\ sample of 19~DLAs therefore defines a representative set of normal, 
star-forming systems in their early phases of formation.


Our ALMA observations of two DLAs in an initial low-resolution \cplus\ search resulted in 
the discovery of strong \cplus\ emission from galaxies associated with the DLAs \citep{neeleman+17}. In one of these 
galaxies, at $z \approx \zdla$ towards \dlaname, the \cplus\ morphology and kinematics 
showed evidence for a complex system, which we suggested might arise from a pair of 
merging galaxies. Here, we present new ALMA observations, using a more extended 
configuration, aiming to resolve the \cplus\ and dust far-infrared (FIR) continuum 
emission.


This paper is organized as follows. Section~\ref{sec:obs} presents the new ALMA observations, while Section~\ref{sec:discuss}
characterizes the morphology of the merging galaxies, and discusses the implications for estimating \fmg\ at $z>2$. We conclude in Section~\ref{sec:conclude} with thoughts on the future and a brief summary. Throughout the manuscript, we adopt a $\Lambda$-Cold Dark Matter cosmology with cosmological parameters as defined in  \citet{Planck2015}, and report physical distances unless otherwise noted.

\section{Observations and Data Analysis}
\label{sec:obs}

The $z = \zdla$ DLA towards \jqso\ (hereafter \dlaname), with an \hi\ column 
density of $\mnhi = 10^{21.35}~\cmt$, was discovered by a DLA survey of the 
Sloan Digital Sky Survey \citep{phw05}. Follow-up Keck/ESI observations yielded 
a high gas-phase metallicity ([M/H]~$\approx -0.75$) and a large velocity width \dvninety\ 
for its low-ionization metal-line profiles \citep{rafelski+12}. Because of these features, we selected it for our original ALMA observations, which yielded detections of both 
\cplus\ line and dust continuum emission from a galaxy offset by $\approx \vrperp$\,kpc
from the quasar sightline at $z = 3.7978$ \citep{neeleman+17}.

\begin{figure*}
\centering
\includegraphics[width=\textwidth]{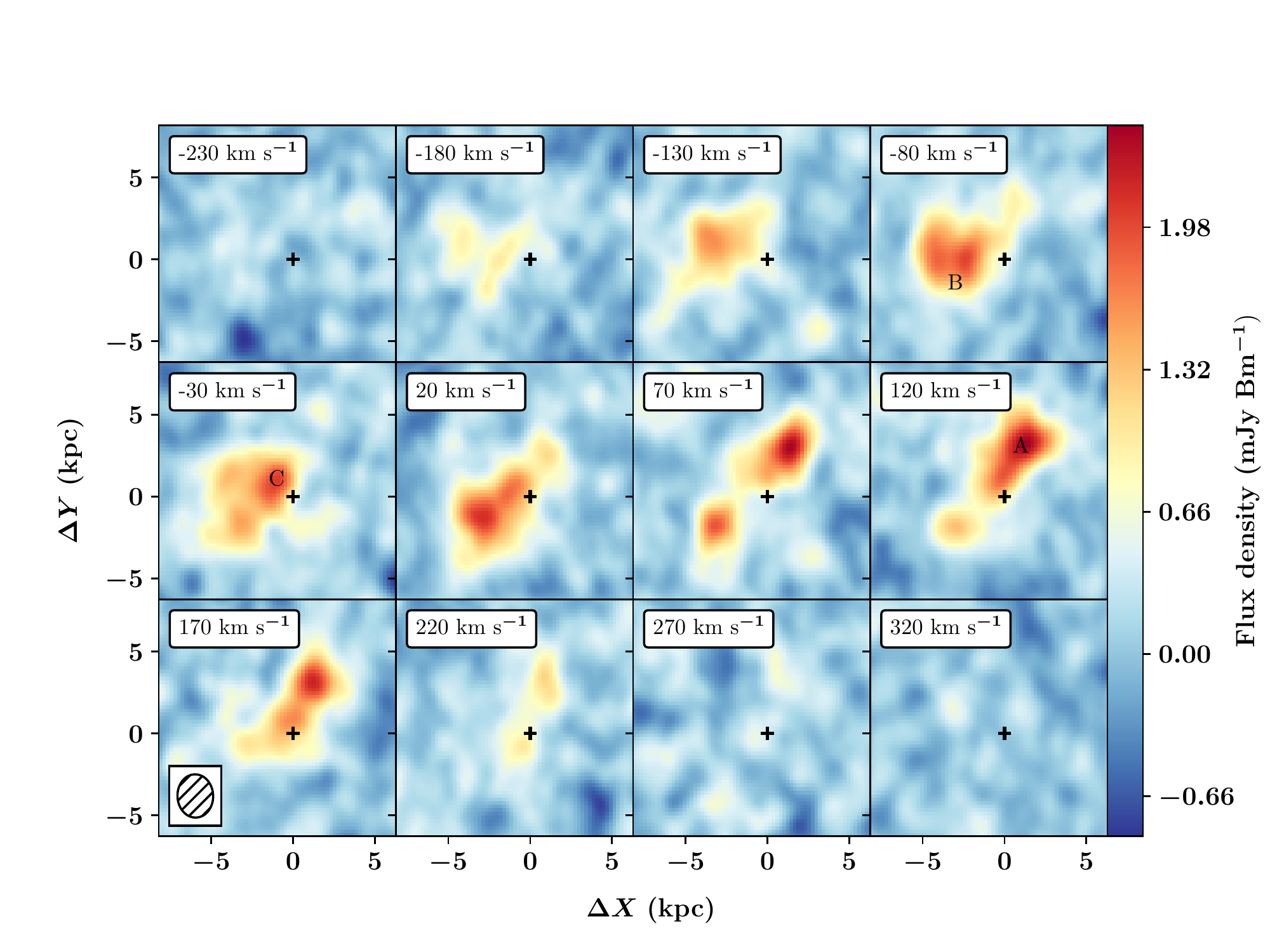}
\vspace{-0.3in}
\caption{Channel maps of the \cplus\ emission from the gas associated with 
\dlaname, relative to the redshift $z = \zcube$. We label these emitting components 
(A, B, and C), as defined in the main text; these have distinct spatial and/or 
kinematic signatures.  Units on the color-bar are Jy~Bm$^{-1}$.
}
\label{fig:channel}
\end{figure*}

Owing to the kinematic complexity of the \cplus\ line-emission already present in the low resolution observations and its relatively 
high flux density, we used the ALMA Band-7 receivers to observe the $z = 3.7978$ 
galaxy in configuration C43-4, with a higher angular resolution ($\bwidth$, i.e. 
$\approx \spatres$\,kpc at $z=3.7978$), in Cycle~\newcyc, for a total on-source time 
of 2.25~h. We used four 1.875~GHz IF bands for the observations, one centered on the 
redshifted \cplus\ line frequency (at 396.127~GHz), with 3840 channels, and 
the other three at neighbouring frequencies, each with 128~channels to sample the dust continuum of the galaxy. 

The initial data analysis was carried out with the ALMA pipeline in the Common Astronomy Software Application ({\sc casa}) package; this included the removal of bad data (e.g. due to non-working antennas), and calibration of the flux density scale and the antenna-based complex gains and system bandpass. After this, the target source data were split into a separate measurement set, and a self-calibration procedure was carried out on the continuum data, in {\sc casa}, to more accurately measure the antenna-based gains. The self-calibration procedure involved a few rounds of imaging and phase-only self-calibration, followed by a couple of rounds of amplitude-and-phase self-calibration and imaging, until the gain solutions converged and no improvement was seen in the image on further 
self-calibration. The self-calibration used a ``robust'' scheme, with custom 
routines written in the {\sc casa} environment. The final naturally-weighted continuum 
image has a root-mean-square (RMS) noise value of $\approx 30 \mu$Jy/Bm and an angular 
resolution of $0\farcs36 \times 0\farcs31$. The peak signal-to-noise ratio on the image improved from $\approx 100$ before self-calibration to $\approx 330$ after self-calibration, and the image RMS noise improved by a factor of $\approx 1.7$, indicating a significant improvement due to the self-calibration process. The antenna-based gains derived from the above procedure were then applied to the line data. Finally, a spectral cube was made using the visibilities from the band centered on the redshifted \cplus\ frequency, using natural 
weighting. The spectral cube has an RMS noise of $250$~$\mu$Jy/Bm per 50~km/s channel, 
and an angular resolution of $0\farcs37 \times 0\farcs31$.

\begin{figure}
\centering
\includegraphics[width=0.5\textwidth]{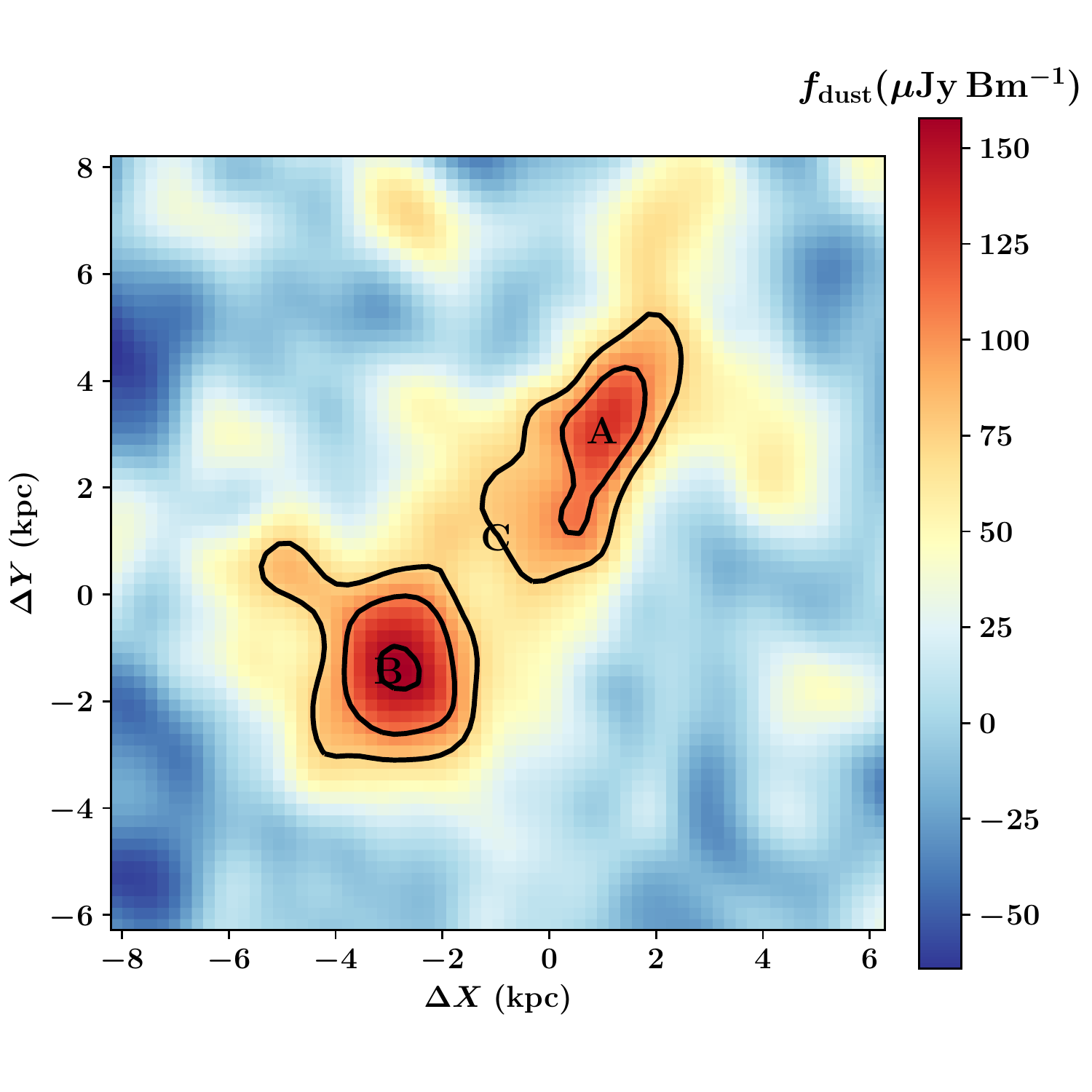}
\vspace{-0.3in}
\caption{The ALMA 403.1~GHz continuum emission from the galaxies associated with DLA1201+2117\_3.80.  Marked are the locations of sources A, B, and C.  
The lowest contour corresponds to $3\sigma$ significance 
(25~$\mu$Jy km s$^{-1}$ beam$^{-1}$), with the contours increasing by a factor 
of 2$^{1/2}$. 
We identify bright emission from
components A and B, but only weak, diffuse emission near C, which may even be predominantly 
from the other two sources.
}
\label{fig:dust}
\end{figure}

\begin{figure}
\centering
\includegraphics[width=0.5\textwidth]{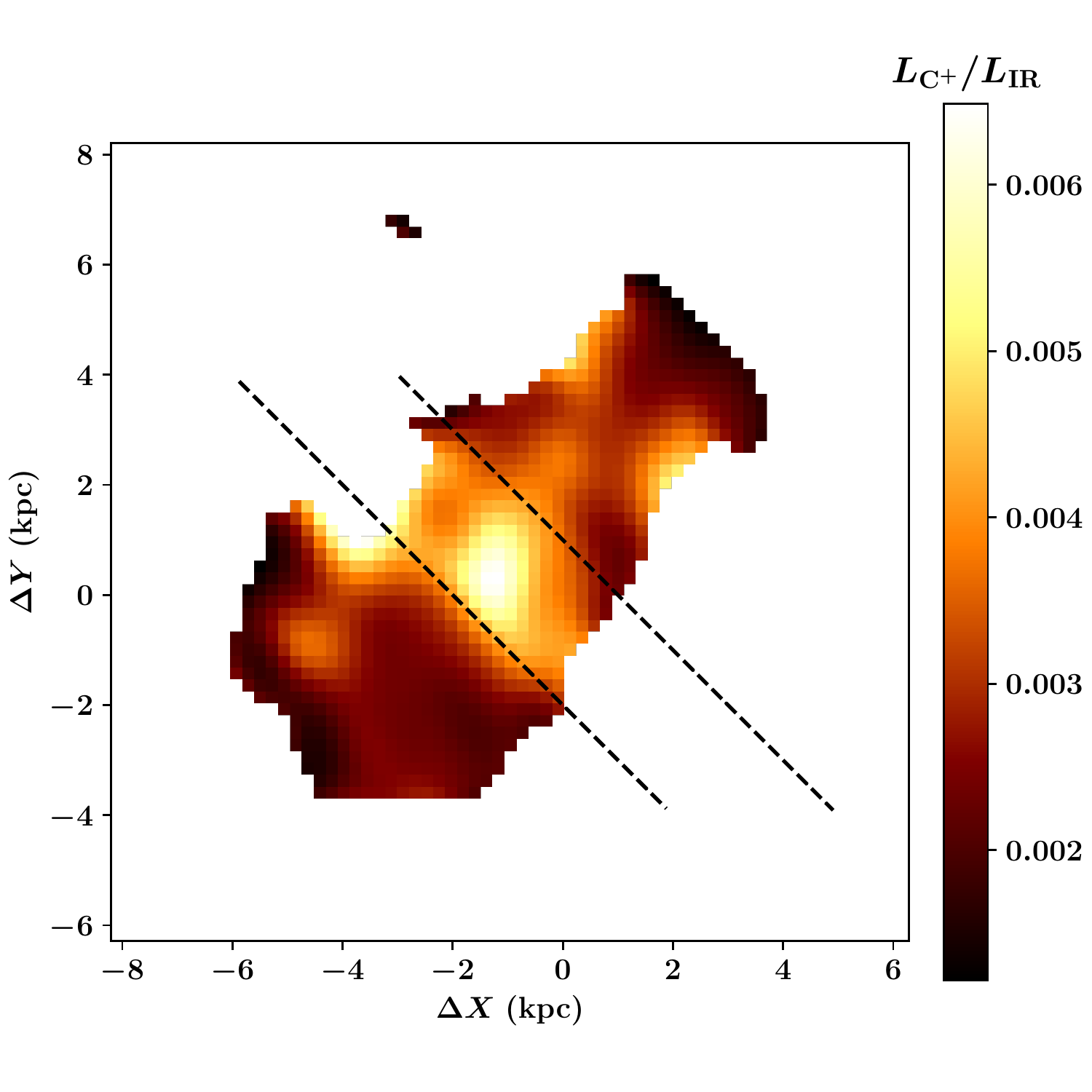}
\vspace{-0.3in}
\caption{Ratio of \lcplus/\ldust\ per spaxel for the gas in the merging system.  
The lines demarcate the A (upper), B (lower) and C (middle) components.  The former 
two have \lcplus/\ldust\ ratios characteristic of star-forming galaxies with modest 
SFRs whereas the ratio for component~C is $\approx 2$ times higher (see Table~\ref{tab:values}).  
Uncertainties in the integrated values are $\approx 1-3 \times 10^{-4}$.
We propose that 
the gas in component~C is stripped material, from one or both of galaxies A and B, and 
the elevated ratio results from \cplus\ emission from predominantly atomic or even ionized gas.
}
\label{fig:ratio}
\end{figure}

Figure~\ref{fig:moments} presents the zeroth and first moments of the \cplus\ emission, 
centered on the galaxy position as determined from the lower resolution data (\origpos). The axes are labeled in physical units (kpc), at $z=3.7978$, with 
$\Delta X$, the negative offset in RA, and $\Delta Y$, the offset in Dec, relative to 
\origpos. At the higher spatial resolution of the new observations, the source reported 
by \citet{neeleman+17} separates into several components, in both space and velocity. 
This is clearly visible in the \cplus\ channel maps (relative to $z=3.7978$) shown in 
Figure~\ref{fig:channel}, which further discriminate between the different components. 
At velocities $\delta v < -50 \, \mkms$, the emission is 
dominated by an extended source located $\approx 3$\,kpc East of \origpos. We refer to 
this source as component~B, and note that it also exhibits substantial emission at $\delta v > 0 \, \mkms$.
At $\delta v > 50 \, \mkms$, the brightest emission occurs 
at $\Delta X \approx +1$\,kpc and $\Delta Y = +3$\,kpc, which we refer to as component~A. 
In between these two components is emission that spans a range of velocities, hereafter 
component~C. For quantitative measurements, we define the following three regions, to 
clearly separate the three components:

\begin{itemize}
  \item A: Emission below the line $\Delta Y = \aslope \,\Delta X \rm \ayoff$
  \item B: Emission above the line $\Delta Y = \bslope \,\Delta X \rm \byoff$
  \item C: emission between the above two lines 
\end{itemize}
These three components are labeled in Figure \ref{fig:moments}. The brightest components (A and B) are also clearly detected as separate dust peaks in the continuum emission map, whereas component C is observed as weak extended emission in the dust continuum (Fig. \ref{fig:dust}). This suggests a scenario of two merging, star-forming galaxies (A and B), with stripped gas dominating the emission in component C. The lines demarcating the different components are shown in Figure~\ref{fig:ratio}.

To test the assertion of multiple sources, we attempted to fit the line emission with 
a single rotating-disk model using the fitting code described in \cite{neeleman+20}. 
The code compares the observed spectral cube to a cube generated from the model convolved with the instrument resolution. The best-fit parameters and uncertainties are then estimated through a Markov Chain Monte 
Carlo analysis of the parameter space. While the first-moment map shows a significant 
velocity shear across the system, the residuals of the models are significantly higher 
than the RMS uncertainty of the data cube. In other words, we find that no single-disk 
model can reliable reproduce the observed \cplus\ line-emission.

%

We therefore conclude that the emission arises from at least two 
physically-distinct galaxies with projected separation $\delta \mrperp \approx \rsep$\,kpc 
and a line-of-sight velocity separation of $\delta v \approx 200 \, \mkms$. Although we cannot 
strictly rule out a chance projection, the probability that the true separation $\delta r$ 
greatly exceeds $\delta \mrperp$ scales as the cube of the ratio of the separations. Thus, 
the probability that the true separation is $> 30$\,kpc is $< 1$\%. The presence of substantial 
\cplus\ emission between the two galaxies (component~C) is further evidence that the galaxies 
are physically nearby. Simulations of merging galaxies report a high probability for galaxies 
with such small separations to undergo a merger \citep[e.g.][]{patton+00}. We therefore 
conclude that the \cplus\ emission data indicate that A and B are in the process of merging.

\section{Discussion}
\label{sec:discuss}

Our discovery of a pair of merging \hi-selected galaxies with ALMA offers a unique opportunity to 
examine the astrophysics of galaxy mergers in the young Universe, and inspires a new assessment 
of the fraction of major mergers. We discuss each of these in turn.

\subsection{The nature of a $z \sim 4$ merger}

In the previous section, we argued that the \cplus\ line-emission associated with \dlaname\ 
arises from a complex of merging galaxies. This is based on the morphology and kinematics
seen in Figures~\ref{fig:moments}, \ref{fig:channel}, and \ref{fig:dust}.  We now describe the 
properties of these components to provide a quantitative assessment of the system that may 
be compared against models of galaxy mergers. Table~\ref{tab:values} lists integrated and
flux-weighted measurements for the line and continuum emission of the components.  For these 
measurements, we have analyzed our naturally-weighted data cube, using simple box-car integrations 
to estimate the integrated flux densities. We have converted the continuum flux density measurements
(obtained at $\nu_{\rm rest} \approx 1.909 \times 10^{12}$\,Hz) to an IR luminosity (spanning 8 to 1,000 
microns) assuming the dust emits as a modified blackbody with $T = 35$\,K and a Raleigh-Jeans slope of $\beta = 1.6$. For a discussion of the uncertainties of these assumptions for this galaxy see \citet{neeleman+17}.

\begin{deluxetable*}{cccccccccc}
\tablecaption{MEASUREMENTS\label{tab:values}}
\tabletypesize{\footnotesize}
\tablehead{\colhead{Quantity}  & \colhead{Unit} 
& \colhead{Value$_{\rm A}^a$} & \colhead{Error$_{\rm A}$} 
& \colhead{Value$_{\rm B}^b$} & \colhead{Error$_{\rm B}$} 
& \colhead{Value$_{\rm C}^c$} & \colhead{Error$_{\rm C}$} 
& \colhead{Value$_{\rm T}^t$} & \colhead{Error$_{\rm T}$} 
\\ 
 } 
\startdata 
$\int S_{\rm [CII]} \, dv$ & Jy km s$^{-1}$& 0.51 & 0.007& 0.51 & 0.006& 0.34 & 0.005& 1.36 & 0.010\\ 
\lcplus & $10^8 \, L_\odot$& 2.5 & 0.0& 2.5 & 0.0& 1.7 & 0.0& 6.6 & 0.1\\ 
$f_{\rm IR}$ & mJy& 0.133 & 0.003& 0.097 & 0.003& 0.039 & 0.002& 0.269 & 0.005\\ 
$L_{\rm IR}^d$ & $10^{10} L_\odot$& 10.0 & 0.2& 7.2 & 0.2& 2.9 & 0.2& 20.1 & 0.4\\ 
SFR & M$_\odot$ yr$^{-1}$& 10.8 & 0.3& 7.8 & 0.2& 3.1 & 0.2& 21.8 & 0.4\\ 
\lcplus / $L_{\rm IR}$ & $10^{-3}$& 2.5 & 0.1& 3.4 & 0.1& 5.8 & 0.3& 3.3 & 0.1\\ 
$v$ & km/s& 44.4 & N/A& 1.7 & N/A& 7.7 & N/A& 19.3 & N/A\\ 
\hline 
\enddata 
\tablenotetext{a}{Component A, defined relative to \origpos\ as
the gas below the line $\rm DEC_{\rm off} = \aslope RA_{\rm off} \ayoff$}
\tablenotetext{b}{Component B, defined relative to \origpos\ as
the gas above the line $\rm DEC_{\rm off} = \bslope RA_{\rm off} \byoff$}
\tablenotetext{c}{Component C, defined as the gas between the lines defining A and B.}
\tablenotetext{t}{Total}
\tablenotetext{d}{IR luminosity for $\lambda = 8$ to 1000 microns, estimated from the continuum flux assuming a modified blackbody with $T=35$\,K.}
\tablecomments{Note that fluxes and quantities derived from them (e.g. SFR) have a systematic error of 15\%.}
\end{deluxetable*} 


%
From the measurements in Table~\ref{tab:values}, we estimate the C$^+$/IR luminosity ratio 
\lcplus/\ldust\ which has been observed to anti-correlate with the IR surface
density $\Sigma_{\rm IR}$ \citep[e.g.][]{smith+17,litke+19} yielding a \cplus\ 
deficit in the most intensely star-forming regions. The \lcplus/\ldust\ values for 
components~A and B are consistent with the values derived from $z>2$ galaxies with 
modest SFRs \citep[e.g.,][]{capak+15}.
This supports the association of these components with individual, normal galaxies.

The integrated flux of central component C, however, shows a much higher \lcplus/\ldust\ ratio: 
a factor of approximately two times higher than the integrated values for the other two 
components. This is illustrated in Figure~\ref{fig:ratio} and Table~\ref{tab:values}. 
Referring to Figure~\ref{fig:moments}, 
the \cplus\ emission in this region is not substantially higher than that from components~A or 
B; the high \lcplus/\ldust\ ratio here is thus driven by low dust continuum emission (Fig.~\ref{fig:dust}) and therefore a low \ldust\ value.  Indeed, the 
ratio in component~C lies close to an extrapolation of the relation between \lcplus/\ldust\ 
and  $\Sigma_{\rm IR}$ to lower $\Sigma_{\rm IR}$ values \citep[see also][]{smith+17}. 
This would imply either a dust-poor environment or, less likely, dust with a 
substantially lower temperature. It further suggests that the \cplus\ emission in 
component~C does not mainly arise from photo-dissociation regions but instead tracks 
primarily atomic or even ionized gas.

Elevated \cplus\ luminosities relative to the dust emission have been reported for Lyman-break
galaxies at $z \sim 5$, possibly due to lower internal dust-to-gas ratios \citep[e.g.][]{capak+15}.
The $z = 3.7978$ galaxy, however, marks the first report of spatial variations within 
a system to such high \lcplus/\ldust\ values. We propose that the gas in component~C is 
material stripped from one (or both) of galaxies A and B. This leads to the picture of 
two merging, star-forming galaxies that have tidally interacted to  pull gas into the region 
between them. This picture is supported by theoretical results on merging galaxies that 
predict significant tidal interactions when galaxies lie within $\approx 10$\,kpc 
\citep{chill+10} 
although a quantitative comparison is beyond the scope of this manuscript.

It is commonly assumed that merger events drive an enhanced SFR  in gas-rich galaxies. 
This assertion has theoretical underpinnings \citep{mh96} and observational support, 
e.g., the frequent detection of multiple components in the most luminous sub-mm galaxies 
\citep{hayward+18}. However, as noted above, components A and B have modest \ldust\ 
values and correspondingly modest SFRs: (A) $\approx 10.8 \pm 1.5 \mmsun \, \, \rm yr^{-1}$ 
and (B) $\approx 7.8 \pm 1 \mmsun \, \rm yr^{-1}$. These values are consistent with 
the local correlation between SFR and \cplus\ line luminosity \citep[][]{delooze11,delooze14}. 
We thus find no evidence for elevated star formation in this merger. Given that it was 
selected in absorption, we speculate that the elevated SFRs reported for other mergers 
may represent a selection bias.


\begin{figure}
\centering
\includegraphics[width=0.5\textwidth]{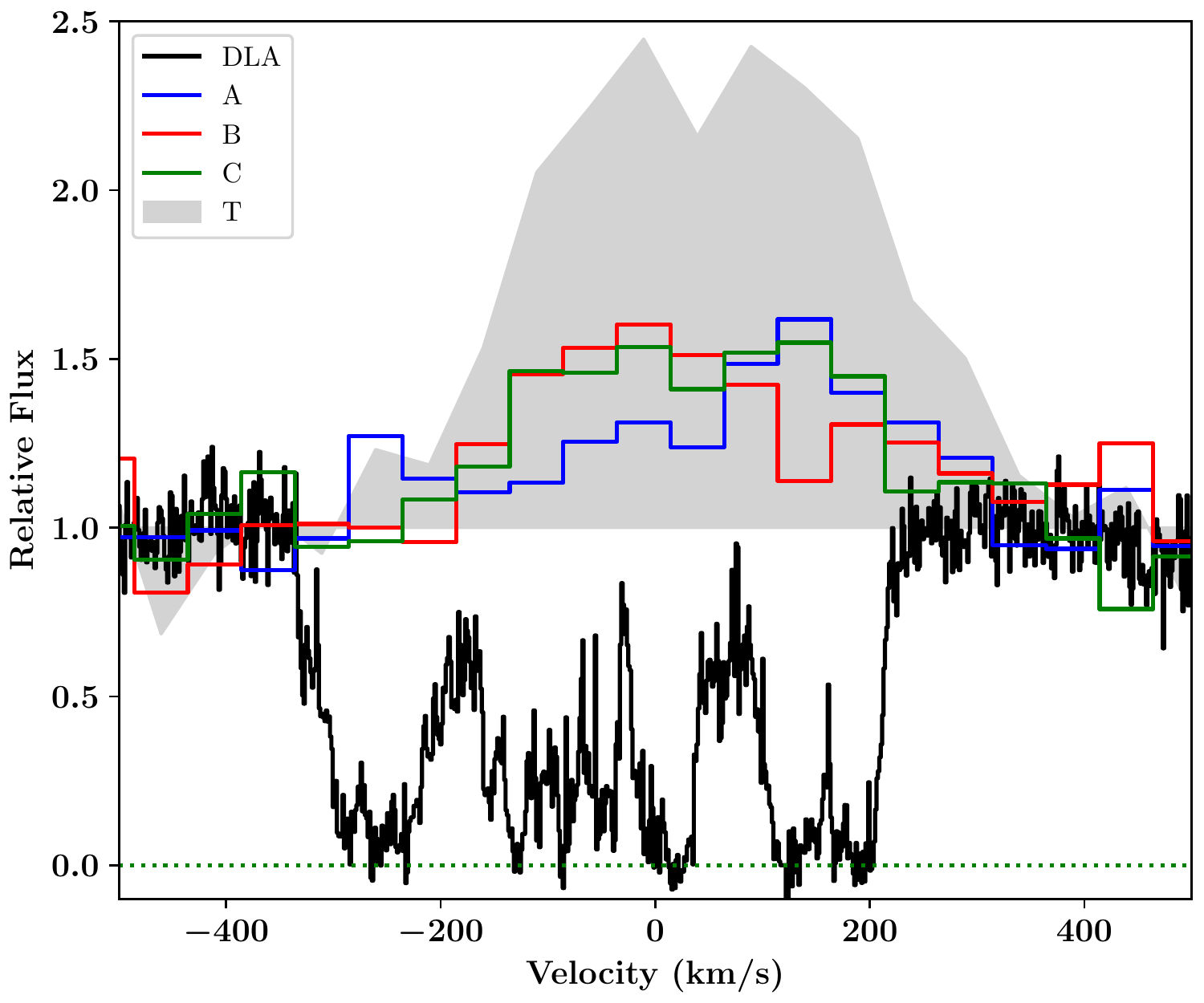}
\vspace{-0.3in}
\caption{Comparison of the kinematic profiles (relative to $z=\zdla$) of the DLA absorption 
(black; the \fetwo~1608 profile) with the \cplus\ emission of galaxies A$-$C (blue, red, green) 
and the total, flux-weighted profile (gray). The galaxy emission kinematics are consistent 
with two of the three main complexes of the DLA, and one is motivated to invoke a causal 
connection, i.e.\ that the galaxy merger yields the large \dvninety\ of the DLA.
}
\label{fig:kinematics}
\end{figure}

\subsection{On the origin of high \dvninety\ DLAs}
\label{sec:high_dv}

Figure~\ref{fig:kinematics} compares the line-of-sight velocity profiles of the three 
\cplus\ components with a low-ionization metal-line absorption profile, each relative 
to $z = \zdla$. The velocity width of the metal-line absorption is extreme, 
$\mdvninety^{\rm DLA} \approx 470 \, \mkms$,  which we again emphasize was part of our selection
criteria for the original ALMA search for \cplus\ emission \citep{neeleman+17}. Tellingly, 
the velocity profiles of the three \cplus\ components span nearly the entire velocity 
width of \dlaname. Specifically, these components correspond closely to the main complexes
of the absorption profile. The only (possible) exception is the absorption at 
$\delta v < -200 \, \mkms$, which corresponds to $\approx 20\%$ of the total optical depth.

\begin{figure}
\centering
\includegraphics[width=0.5\textwidth]{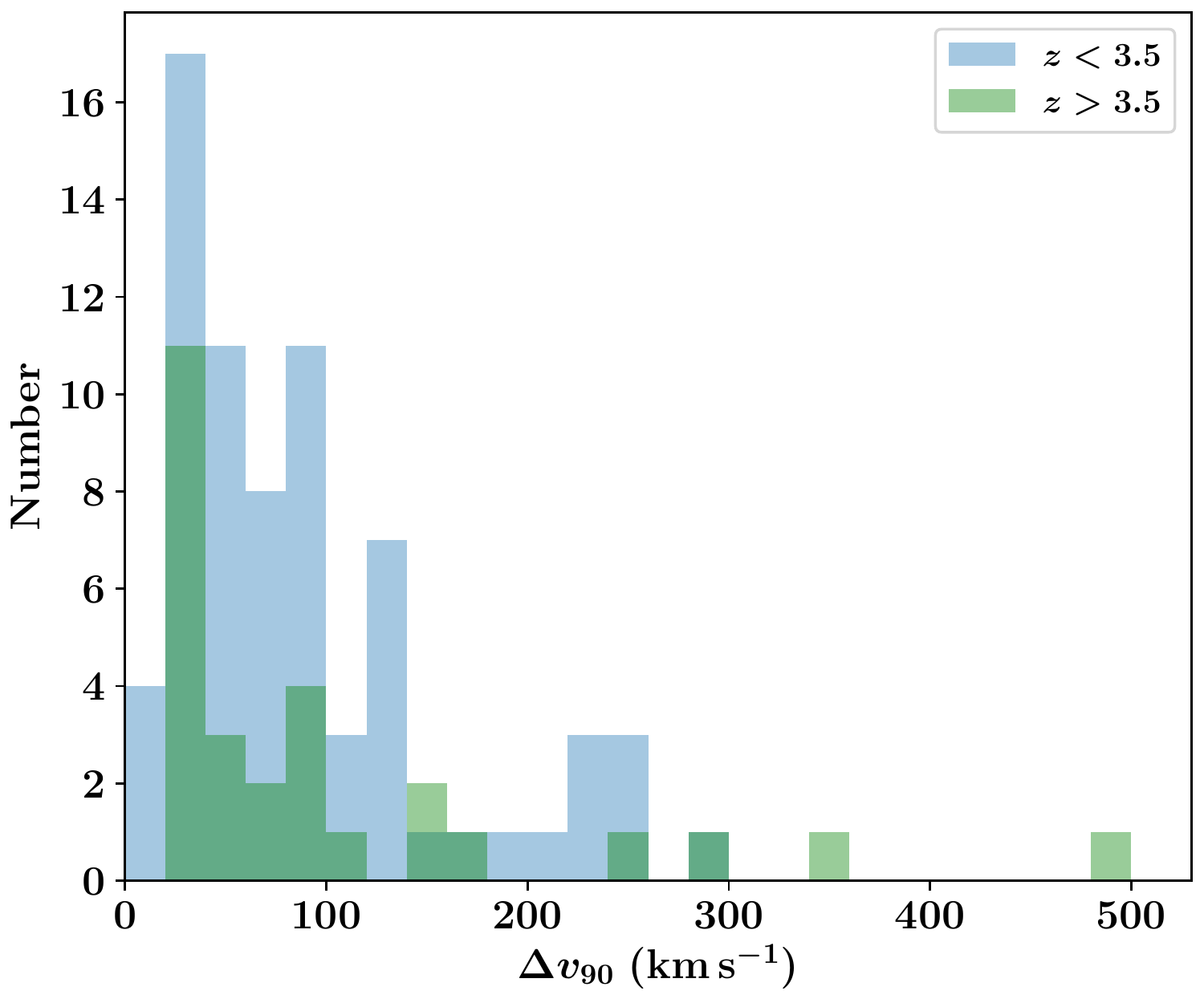}
\vspace{-0.3in}
\caption{Distributions of velocity widths \fdv\ for a sample of $z>2$ DLAs derived without 
kinematic pre-selection, split at $z_{\rm DLA} = 3.5$. Each sub-sample exhibits a median 
\dvninety\ of  $\approx 70-80 \, \mkms$ and a tail extending to $\mdvninety > 200 \, \mkms$.
}
\label{fig:fdv}
\end{figure}

The close correspondence in kinematics between the galaxy's \cplus\ emission and the 
metal-line absorption suggests a causal relationship, i.e.\ that gas associated with 
the merging galaxies contributes to the DLA profile to yield its high \dvninety. 
Indeed, a number of previous studies on DLA kinematics have asserted that the largest values of the
\dvninety\ distribution \fdv\ correspond to DLAs arising from multiple, merging galaxies
\citep[e.g.][]{pw97,mps+01,pgp+08,bird+15}.

Figure~\ref{fig:fdv} shows the \fdv\ distributions measured from high-resolution absorption
studies of DLAs, and is restricted to systems without kinematic pre-selection 
\citep{marcel13,rafelski+12}. The median of each \fdv\ distribution lies at 
$\approx 80 \, \mkms$, which is predicted to reflect the gravitational potential of 
the dark matter halos hosting DLAs \citep{bird+15,pma+18}. One notes further that the 
\fdv\ distribution is highly asymmetric, i.e.\ skewed with a tail to high \dvninety\ values.
The 90th percentile of the $z \sim 2$ and $z \sim 4$ distributions are 
$\approx \dvntwo \, \mkms$ and $\approx \dvnfour \, \mkms$, respectively. These exceed 
the differential velocity that can be produced by a rotating disk, even with extreme 
thickness, inclination, and rotation speed \citep{pw97}. This has motivated interpretions 
involving multiple galaxies and mergers, to account for the high \dvninety\ values 
\citep[and the occasional call for non-gravitational motions, e.g.\ winds;][]{nbf98}. 

While \dlaname\ offers only a single example, the results described here support 
the merger interpretation for DLAs with the highest \dvninety\ values. Evidence for 
companion galaxies at the DLA redshift was also obtained by \citet{neeleman+19} in 
another of our \cplus\ emitters, at $z \approx 4.224$ towards PSS~1443+2741, albeit 
with a slightly larger separation, $\approx 12$~kpc. This absorber too has a large 
\dvninety\ value, $\approx 284 \, \mkms$ \citep{neeleman+19}.  Adopting this 
interpretation, it provides an explanation for why high-\dvninety\ DLAs show a 
larger dispersion around the correlation between metallicity and DLA kinematics 
\citep{pcw+08}. It also tempers the need to introduce galactic-scale winds that 
accelerate neutral gas to speeds greatly exceeding the escape velocity.

\subsection{Inferring the Major Merger Fraction}
\label{sec:mrate}

Having associated large \dvninety\ values from DLAs with galaxy mergers, we proceed 
to use DLA statistics to roughly infer the (major) merger fraction of galaxies at $z>2$. 
We make the ansatz that every DLA with $\mdvninety > \mdvmax$ tags an ongoing major 
merger. The ratio of the number of DLAs with $\mdvninety > \mdvmax$ to the total 
number of DLAs then immediately yields the fraction of DLAs undergoing a merger.

\begin{figure}
\centering
\includegraphics[width=0.5\textwidth]{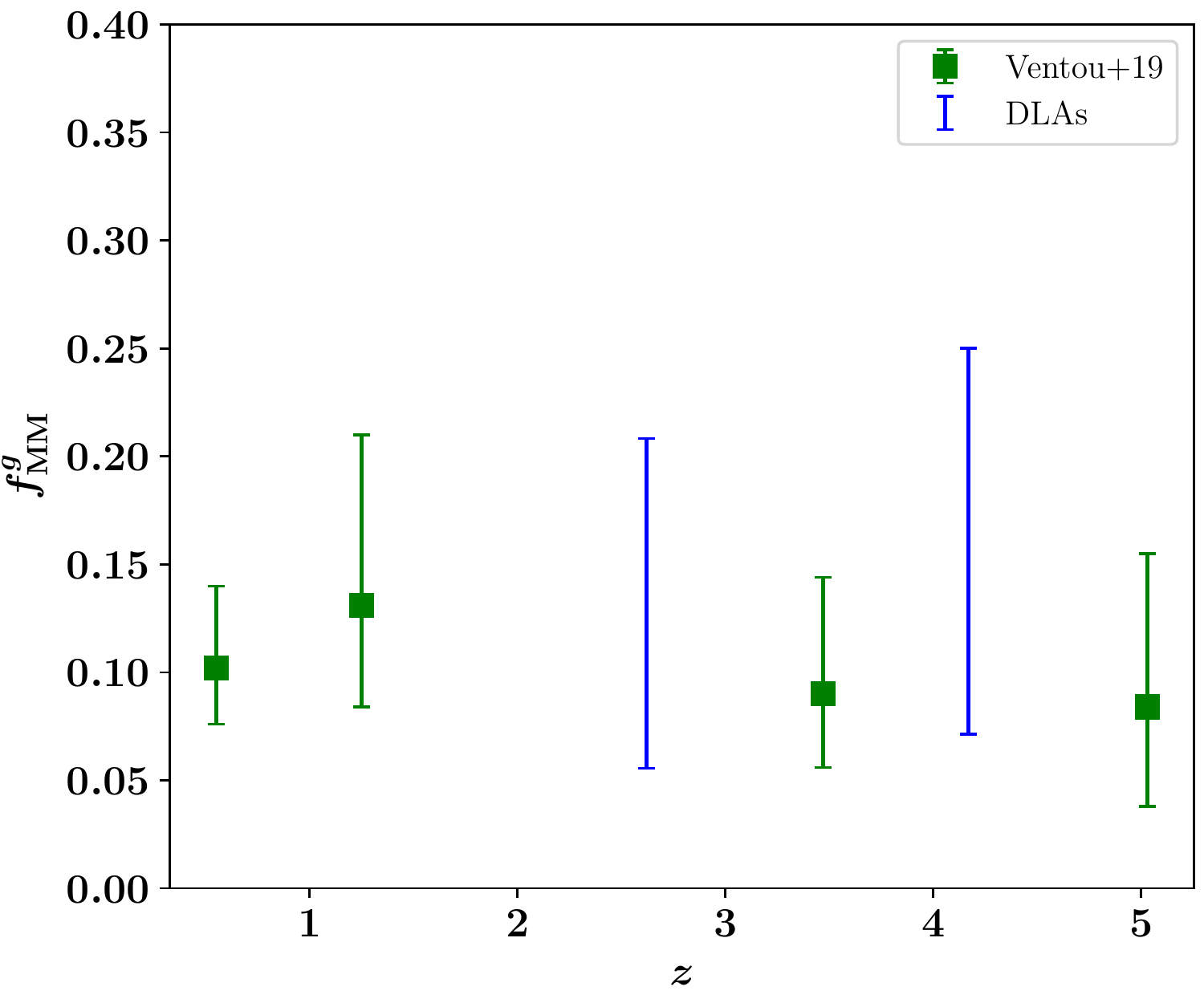}
\vspace{-0.3in}
\caption{Estimates for the fraction of galaxies undergoing a major merger as
a function of redshift.  Green squares are from \cite{ventou+2019} 
for $9.5 < \log(M_*) < 11$; these authors used spectroscopically-confirmed galaxies
to estimate \fmg. The blue bars are estimates derived here from the incidence of DLAs
with high low-ionization metal line velocity widths, \dvninety.  These estimates 
includes uncertainties from several systematics and underlying assumptions (see main text 
for discussion).
}
\label{fig:fmm}
\end{figure}

\begin{equation}
    \mfmdla = \frac{\intl_{\mdvmax}^\infty \mfdv \, d\mdvninety}{
    \intl_0^\infty \mfdv \, d\mdvninety}
\end{equation}
Taking $\mdvmax = 150 \, \mkms$, as inferred from the \fdv\ distributions in 
Fig.~\ref{fig:fdv}, 
we estimate $\mfmdla = 0.14$ and $0.18$ for $z \sim 2$ and 4, respectively. 
Given that there would not be a sharp transition in \dvmax\ on the occurence of 
a merger, we allow \dvmax\ to range from $125$ to $200 \, \mkms$, and thus estimate 
the systematic uncertainty on these \fmdla\ estimates.

Orientation and projection effects may yield $\mdvninety < \mdvmax$ for some mergers;
one might hence adopt these \fmdla\ values as lower limits to the DLA merger 
fraction.  Conversely, it is possible that other physical processes contribute 
to large \dvninety\ values, implying that the \fmdla\ values are over-estimates.
This could include minor mergers, although numerical simulations suggest that their 
contributions are small \citep{bird+15}. We proceed with our nominal approach and 
emphasize that this yields only a coarse estimate of the major merger fraction. Analysis 
of simulations could constrain these effects; however, this is beyond the scope of this paper.

To roughly estimate the merger fraction of galaxies, one must account for the 
cross-sectional weighting that is inherent to DLA analysis, i.e.\ systems with 
a larger projected area \adla\ whose \hi\ column density exceeds the DLA threshold 
have a higher probability of intersecting a quasar sightline. If the merger does 
not affect \adla, then we expect the merging system to have $\madla^{\rm merger} = 
2 \madla$, i.e.\ twice the area owing to the presence of two galaxies. 
This leads to an estimate for the fraction of galaxies undergoing a major 
merger: $\mfmg = \mfmdla / (\madla^{\rm merger} / 2 \madla)$.
Any enhancement due to the merger, e.g.\ from stripped gas as inferred for the 
system presented here, would require a downward correction. In the following, we 
include a systematic uncertainty to $\mfmg$, corresponding to  allowing 
$\madla^{\rm merger}$ to range from 2 to 4 $\times$ \adla.

%

Figure~\ref{fig:fmm} presents our first estimates of \fmg\ from DLA surveys (blue).  
These 
are presented as error intervals that express the large systematic uncertainties described 
above. They are compared with estimates of \fmg\ from the incidence of 
spectroscopically-confirmed galaxies with $\delta \mrperp < 25$\,kpc and 
$\delta v < 500 \, \mkms$ \citep[green; ][]{ventou+2017}. It is clear that our 
results are consistent with the directly observed merger fraction \citep{ventou+2019}, 
albeit with large systematic uncertainties. We do not observe the decrease in $\mfmg$ 
at higher redshifts, although our large systematics do not allow us to constrain the 
evolution with redshift.

\section{Summary}
\label{sec:conclude}

We have used the ALMA Band-7 receivers to image, at $\approx 2$~kpc resolution, the \cplus\ line 
and FIR dust-continuum emission from an absorption-selected galaxy at $z = \zcube$. 
The \cplus\ and dust emission arise from at least three distinct source components 
lying within $\approx 10$\,kpc (projected), two of which (components A and B) have 
\lcplus/\ldust\ ratios consistent with values
in high-$z$ galaxies with moderate SFRs, while the third (component C), 
has an elevated \lcplus/\ldust\ 
ratio. The images are consistent with a scenario in which components A and B are 
normal star-forming 
galaxies separated by $\approx \rsep$\,kpc and undergoing a major merger, while 
component~C arises from 
gas that has been stripped from one or both of the galaxies. 
The velocity spread of the \cplus\ emission 
is in good agreement with the large velocity spread of the low-ionization metal absorption 
lines, suggesting that the merger is likely to have caused the large \dvninety\ value of the DLA. 

This is the first clear evidence that some of the large \dvninety\ values observed in DLAs arises from merging
galaxies. We propose that major mergers at high redshifts can be identified in a luminosity-independent
manner from the fraction of DLAs with large \dvninety\ values. We use the \dvninety\ distribution 
of DLAs at $z > 3$ to roughly estimate the fraction of major mergers 
amongst normal high-$z$ galaxies $\mfmg$, 
obtaining $\mfmg \approx 0.1$ at $z\approx 2.5$ and $z \approx 4$, 
consistent with independent estimates 
from recent VLT/MUSE studies (albeit with large systematic uncertainty).

\acknowledgments
This paper makes use of the following ALMA data: ADS/JAO.ALMA \#2017.1.01052.S. ALMA 
is a partnership of ESO (representing its member states), NSF (USA) and NINS (Japan), 
together with NRC (Canada), MOST and ASIAA (Taiwan), and KASI (Republic of Korea), in 
cooperation with the Republic of Chile. The Joint ALMA Observatory is operated by ESO, 
AUI/NRAO and NAOJ. M.N. acknowledges support from ERC Advanced Grant 740246 (Cosmic{\verb|_|}Gas). 
NK acknowledges support from the Department of Science and Technology via a Swarnajayanti 
Fellowship (DST/SJF/PSA-01/2012-13). NK thanks Aditya Chowdhury for discussion and the use of 
his robust calibration routines for the analysis.


\bibliographystyle{apj}
\bibliography{alma_merger}

\end{document}